\documentclass[showpacs,twocolumn,aps,superscriptaddress,floatfix,final,prx]{revtex4}
\usepackage{mathptmx}
\usepackage{color}

\usepackage[latin9]{inputenc}
\usepackage{float}
\usepackage{amsmath}
\usepackage{amssymb}
\usepackage{graphicx}

\makeatletter

\DeclareRobustCommand{\greektext}{%
  \fontencoding{LGR}\selectfont\def\encodingdefault{LGR}}
\DeclareRobustCommand{\textgreek}[1]{\leavevmode{%
  \IfFileExists{grtm10.tfm}{}{\fontfamily{cmr}}\greektext #1}}
\DeclareFontEncoding{LGR}{}{}
\DeclareTextSymbol{\~}{LGR}{126}

\@ifundefined{textcolor}{}
{%
 \definecolor{BLACK}{gray}{0}
 \definecolor{WHITE}{gray}{1}
 \definecolor{RED}{rgb}{1,0,0}
 \definecolor{GREEN}{rgb}{0,1,0}
 \definecolor{BLUE}{rgb}{0,0,1}
 \definecolor{CYAN}{cmyk}{1,0,0,0}
 \definecolor{MAGENTA}{cmyk}{0,1,0,0}
 \definecolor{YELLOW}{cmyk}{0,0,1,0}
}

\usepackage{dcolumn}\usepackage{bm}\usepackage{amsfonts}\usepackage{epsfig}\usepackage{dcolumn}\usepackage{bm}

\makeatother

\begin{document}

\title{Shock-waves and commutation speed of memristors}

\author{Shao Tang}

\affiliation{Department of Physics and National High Magnetic Field Laboratory,
Florida State University, Tallahassee, FL 32306, USA}

\author{Federico Tesler}

\affiliation{Departamento de F\'{i}sica - IFIBA, FCEN, Universidad de Buenos Aires, Ciudad
Universitaria Pabell\'on I, (1428) Buenos Aires, Argentina}

\author{Fernando Gomez Marlasca}

\affiliation{GIA-CAC - CNEA, Av. Gral Paz 1499 (1650) San Mart\'{i}n, Pcia Buenos Aires, Argentina.}

\author{Pablo Levy}

\affiliation{GIA-CAC - CNEA, Av. Gral Paz 1499 (1650) San Mart\'{i}n, Pcia Buenos Aires, Argentina.}

\author{V. Dobrosavljevi\'{c}}

\affiliation{Department of Physics and National High Magnetic Field Laboratory,
Florida State University, Tallahassee, FL 32306, USA}

\author{Marcelo Rozenberg}

\affiliation{Laboratoire de Physique des Solides, CNRS-UMR8502, Universit\'e de
Paris-Sud, Orsay 91405, France}

\begin{abstract}
Progress of silicon based technology is nearing its physical limit, as minimum feature size 
of components is reaching a mere 10 nm.
The resistive switching behaviour of transition metal oxides and the associated memristor 
device is emerging as a competitive technology for next generation electronics. 
Significant progress has already been made in the past decade and devices are beginning 
to hit the market; however, it has been mainly the result of empirical trial and error.
Hence, gaining theoretical insight is of essence. In the present work we report the
striking result of a connection between the resistive switching and {\em shock wave} formation,
a classic topic of non-linear dynamics.
We argue that the profile of oxygen vacancies that migrate during the commutation 
forms a shock wave that propagates through a highly resistive region of the device.
We  validate the scenario by means of model simulations and experiments in a 
manganese-oxide 
based memristor device. 
The shock wave scenario brings unprecedented physical insight and enables to rationalize 
the process of oxygen-vacancy-driven resistive change with direct implications for a key technological aspect -- the 
commutation speed.
\end{abstract}


\maketitle

The information age we live in is made possible by a physical underlayer of electronic hardware,
which originates in condensed matter physics research. Despite the mighty  
progress made in recent decades, the demand for faster and power efficient devices
continues to grow.
Thus, there is urgent need to identify novel materials and physical mechanisms for future
electronic device applications. In this context, transition metal oxides (TMOs) are
capturing a great deal of attention for non-volatile memory applications \cite{roadmap}.
In particular, TMO are associated to the phenomenon of resistive 
switching (RS) \cite{scholarpedia} and the memristor device \cite{HP} that is emerging 
as a competitive technology for next generation electronics 
\cite{roadmap,yang2008memristive,waser-aono,waser2009AdvMat,Sawa2008,inoue-sawa,MRS,baikalov2003field}.
The RS effect is a large, rapid, non-volatile, and reversible change of the resistance,
which may be used to encode logic information. In the simplest case one may associate 
high and low resistance values to binary states, but multi-bit memory cells are also possible 
\cite{IEEE,HP1}.

Typical systems where RS is observed are two-terminal capacitor-like devices, 
where the dielectric might be a TMO and the electrodes are ordinary metals. 
The phenomenon occurs in a strikingly large variety of systems. Ranging from simple
binary compounds, such as NiO, TiO$_2$, ZnO, Ta$_2$O$_5$, HfO$_2$ and CuO, to more complex perovskite
structures, such as superconducting cuprates and colossal magnetoresistive manganites
\cite{scholarpedia,waser-aono,Sawa2008,waser2009AdvMat}.

From a conceptual point of view, the main challenges for a non-volatile memory are: 
(i) to change its resistance within nano seconds (required for modern electronics applications),
(ii) to be able to retain the state for years (i.e. non-volatile), and 
(iii) to reliably commute the state hundreds of thousands of times.

Through extensive experimental work in the past decade, a consensus has emerged around the notion that the change in 
resistance is due to migration of ionic species, including oxygen vacancies ($V_O$), across different regions of the device,  
affecting the local transport properties of the oxide. 
In particular, the important role of highly resistive interfaces, 
such as Schottky barriers, has also 
been pointed out \cite{Sawa2008, inoue-sawa,chen}.

In contrast with the experimental efforts,  theoretical studies remain relatively scare. 
A few phenomenological models were proposed and numerically investigated, which captured different aspects 
of the observed effects \cite{prl2004,HP,ielmini,prb}. 

\begin{figure}
\includegraphics[width=0.4\textwidth]{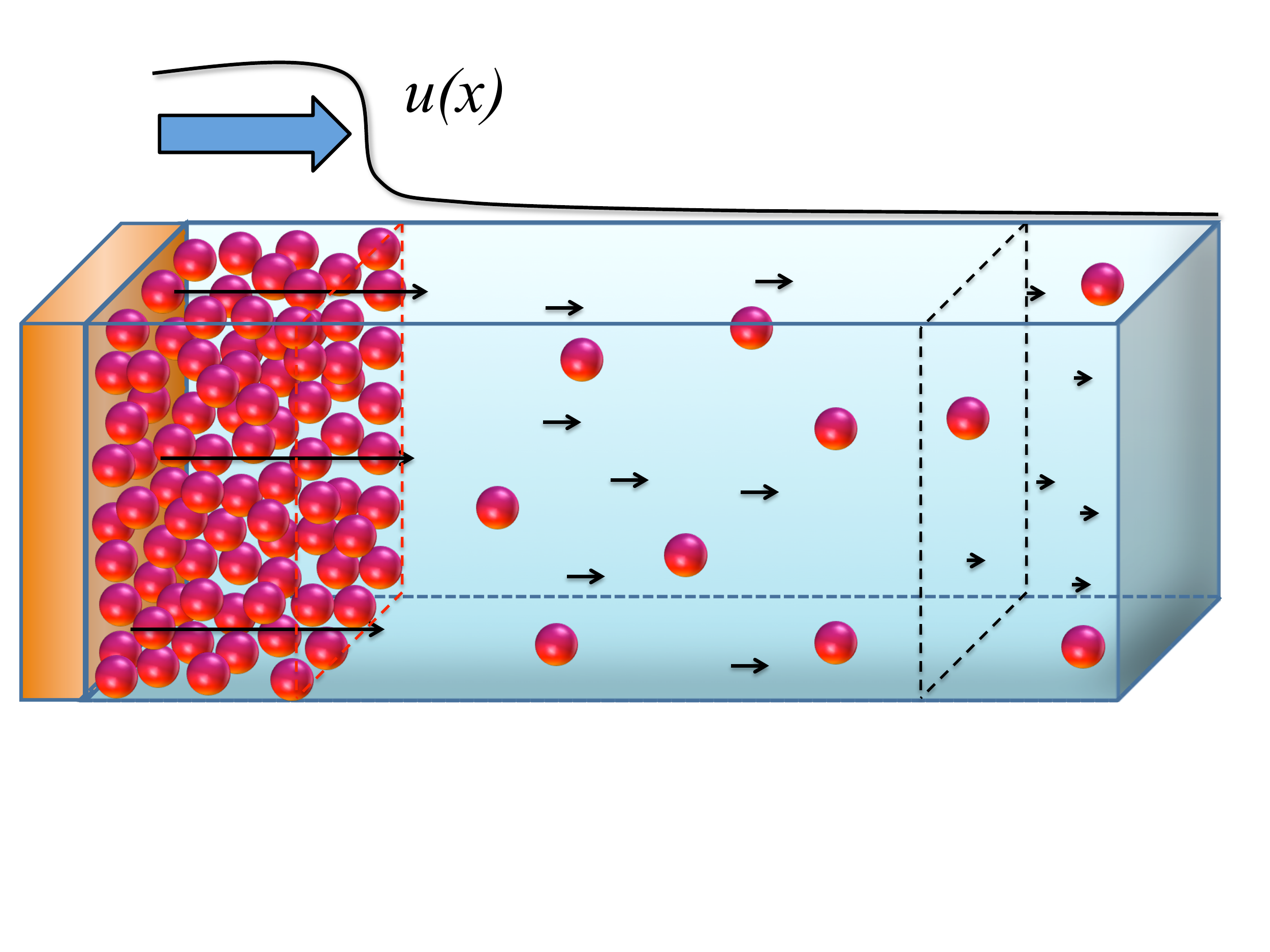}  
\caption{\label{fig:shock} Schematic representation of the shock wave evolution.
The orange region indicates the metallic electrode and the blue indicates the TMO dielectric. Small spheres
denote the ionic defects (oxygen vacancies) whose density profile form a shock wave. It evolves
through a highly resistive (Schottky) interface and eventually leaks over the more conductive bulk, producing
the resistive change. Black arrows depict the strength of the local electric fields.}
\end{figure}

In this context gaining theoretical insight is of essence. Thus, in the present work we shall address one of the
key aspects of the RS phenomenon, namely, the issue of the commutation speed of the resistance change. 
Our first striking result is a connection between the RS phenomenon and {\em shock wave} formation,
a classic topic of non-linear dynamics \cite{NLD}. In fact, we shall argue that the profile of oxygen vacancies that
migrate during the resistive change forms a shock wave that propagates through the Schottky barrier and
leaks onto the bulk of the device, which we schematically illustrate in Fig.\ref{fig:shock}.
We further validate the scenario by means of numerical simulations on a successful model of 
RS and by novel experiments done on a manganese based memristor device. Both model calculations
and experiments reveal a striking scaling behaviour as predicted by the shock wave scenario.

{\em Generalized Burgers' equation.}
When ions migrate through a {\em conducting} medium under the influence
of strong applied voltage, they are likely to undergo a nonlinear
diffusion process, as we explain in the following. The total ionic current
$\mathbf{j}(t,\mathbf{x})=\mathbf{j}_{diffusion}+\mathbf{j}_{drift}$
can be expressed as the sum of a diffusion current $\mathbf{j}_{diffusion}=-D\nabla u$
and a drift current $\mathbf{j}_{drift}$, which is induced by the
local electric field $\mathbf{E}$ and the local concentration
$u$.  Together with the continuity equation $\partial_{t}u+\nabla\cdot\mathbf{j}(\mathbf{x},t)=0$, this
immediately gives us a generalized diffusion equation of the \textit{Nernst\textendash{}Planck
type}. This would represent a linear driven diffusion equation, where the
local electric field $\mathbf{E}$ to be held constant, i.e. independent of the local ion 
concentration $u$. In contrast, in (poorly) conducting media and under voltage pulses, the local electric field may strongly 
depend on the local ion concentration; this effect is the key source of nonlinearity causing 
the formation of shock waves and very sudden resistance switching \cite{inoue-sawa,MRS}. 

Since electrons move much faster than the ions, we can view the ions as
static when considering the electronic current $\mathbf{I}$,
which obeys a steady-state condition $\nabla\cdot\mathbf{I}=0$.
The local electric field is then simply determined, through Ohm\textquoteright{}s
law, by the local resistivity $\mathbf{E}=\rho(u)\mathbf{I}$,
which may be a strong function of the local ion concentration $u$.
In particular, in bad metals such as the transition metal oxides,
the migrating ions (e.g. oxygen vacancies) act as scattering centers
for the conduction electrons. In such situations, we expect $\rho(u)$
to be a \textit{monotonically increasing} function of the local ion
density $u(\mathbf{x},t)$. Therefore, the redistribution
of the local ion density results in the change of local resistivities
and, consequently, of the local electric fields, which further promotes
the non-linear effect in the drift. 

Under the experimentally-relevant case where the transverse currents may be neglected,
the problem simplifies to a one-dimensional
non-linear diffusion equation,
\begin{equation}
\partial_{t}u+f\left(u\right)\partial_{x}u=D\partial_{xx}u,\label{burgers}
\end{equation}
where $f\left(u\right)\equiv\partial_{u}j_{drift}\left(u,I\right)$, and
$I(t)$ is the magnitude of the electronic current.
Equation (\ref{burgers}) can be considered a generalization of the
famous \emph{Burgers' }\textit{\emph{equation}}\emph{, }which corresponds
to the special case $f\left(u\right)\propto u$. Its most significant
feature is the presence of a density-dependent drift term, {\em which physically
means that the ``crest of the wave'' experiences a stronger external
force than the ``trough''}. This generally leads to the formation
of a sharply defined \emph{shock-wave }front in the 
u(x,t) profile, which assumes a universal
form at long times, completely independent of the - quickly ``forgotten''
- initial conditions.  Although the process is driven
by the drift term, the stability
of the shock wave form is provided by the existence of the diffusion term
which prevents the shock wave from self-breaking\cite{taylor2011partial,debnath2011nonlinear}. 
Remarkably, the formation
of shock waves proves to be robust in a much more general family
of models with the nonlinear drift term specified by the function
$f(u)$, any \emph{monotonically increasing} function of
$u$. The qualitative behaviour can be established by using the well-known
``method of characteristics''\cite{debnath2011nonlinear,courant1962methods}, 
as we explain in more detail in the Supplementary Materials.

The drift current is generally given by the expression $j_{drift}=ug(E)$. The
form of the function $g(E)$ is material-dependent, and here we envision
two limiting situations. In homogeneous conductors, we should have
simple ``Ohmic''  behaviour as $g(E)\sim E$
while in granular materials, we expect exponential dependence due to
activated transport, corresponding to: $g(E)\sim\sinh\left(E/E_{0}\right)$, 
where $E_{0}$ is a parameter describing the activation process.

\begin{figure}[t]
\includegraphics[width=0.49\textwidth]{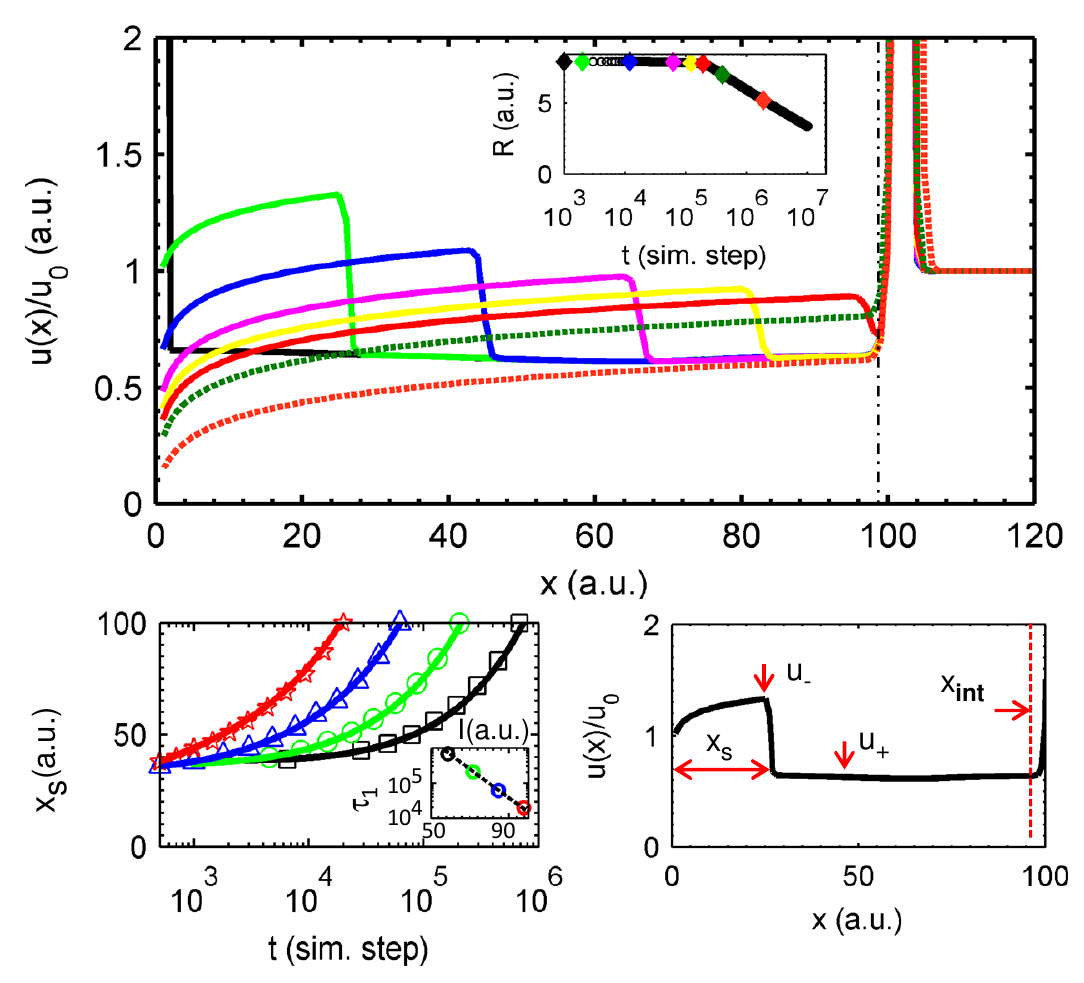}  

\caption{\label{fig:shock wave dynamics} Top panel: Snap shots of the time evolution of the [$V_O(x)$]=$u(x)$ 
profile within the active interfacial region in a simulation of the VEOVM (see Sup. Mat. for details). 
The current $I[a.u]$=11 and $A_{S}$=1000 and $A_{B}$=1.   
The time steps of the successive profiles can be read-off from the corresponding color dots
in the inset. The initial state $u_0(x)$ (black line) exhibits a vacancy pile-up next to the electrode at $x$=1. 
The SB-B interface is denoted with a vertical dash-dot line at $x_{int}$=100.
The large accumulation of vacancies on the right of $x_{int}$ (bulk side) results from 
the initial ``forming'' cycles that conform an approximately fixed background \cite{prb}. 
Inset: Resistance of the device as a function of time. Color dots indicate the value of $R(t)$ 
at the corresponding snapshots of the main panel. 
Bottom-left panel: evolution of the shock wave front position $x_s(t)$ for different 
currents ($I[a.u]$=58.5, 71.5, 84.5 and 97.5). Dots are from numerical simulations and
the solid lines are analytic fits from integration of Eq.~(\ref{eq:shock wave velocity}).  
Inset: characteristic impact-time $\tau_{1}$ as a function of applied currents from the numerical simulation (circles) 
and analytic fit (dotted-line) in semi-log scale. Bottom-right panel: Shock wave parameters.  }
\end{figure}

Remarkably, these general ideas find an explicit realization 
in the context of RS in transition metal oxide memristors,
such as manganites \cite{chen,lee}.
In fact, their transport properties are very sensitively dependent on the oxygen stoichiometry,
i.e. on the concentration of oxygen vacancies [$V_O$].
Thus, it is now widely accepted that the mechanism of the bipolar (i.e. polarity dependent) RS in those
systems is due to the induced changes in the spatial distribution of 
$[V_O({\bf x})] \equiv u({\bf x})$ by means of externally applied strong electric stress \cite{inoue-sawa,MRS}. 
In particular, the accumulation of vacancies within highly resistive regions between the oxide and 
the metallic electrode, such as Schottky barrier (SB) interfaces,
greatly increases the (two-terminal) resistance across the device \cite{prb}. 
This accumulation can be achieved by applying strong voltage pulses across the device, 
leading to the high resistance state $R_{HI}$. Abrupt resistance switching from such  
high-resistance state to a significantly lower resistance state can be accomplished by 
reversing the voltage applied, which removes a significant fraction of vacancies 
from the SB region. The precise characterization of this resistance 
switching process is the main subject of this paper. 

We should mention that an important assumption is 
that the nonlinear drift term plays the dominant role as compared to the normal diffusion, i.e. we shall not
be concerned with the resistive changes involving thermal effects \cite{inoue-sawa,MRS}.
This restriction enables us to apply our analytical tools in a simple manner, allowing us to obtain a simplified mathematical description 
of the migration process, as we show in the following. 

{\em Model system.}
For concreteness, we adopt the \emph{voltage-enhanced oxygen-vacancy migration model} \cite{prb} (VEOVM), which 
corresponds to granular materials with activated transport process and has been
previously used for manganite devices \cite{prb}.
Within the framework of this model, we shall perform numerical simulations to validate our shock-wave scenario. 
The VEOVM simply assumes that the local resistance of the cell at (discretized) position $x$ along the conductive path of the device
is simply given as a linear function of the local vacancy concentration, namely,
\begin{equation}
r(x)=A_{\alpha}u(x)
\label{resistance}
\end{equation}
with $\alpha={S,B}$, where $S$ denotes the highly resistive (Schottky barrier) region and $B$ the 
more conductive bulk \cite{chen}.  
The values of these constants are taken $A_{S}\gg A_{B}=1$, which allows us to neglect the bulk resistance \cite{prb}.  
The discretized conducting assumes the metal-electrode at $x$=0
and $x=x_{int}$ denotes the point within the dielectric where the SB meets the bulk region.  
Under the action of the external stress (electric current $I$), the local fields at each cell position $x$ are computed at every discrete time step $t$. 
The field-driven migration of vacancies is simulated computing the local ionic migration rates from cell $x$ to $x$+$\Delta x$ as \cite{prb}
\begin{equation}
P(x,x+\Delta x) = u(x)[1-u(x)]\exp{\left(\frac{-V_0 + qIr(x)}{k_B T}\right)},
\label{model}
\end{equation}
where, for simplicity, we take the ionic charge $q$=1 and $k_B T$=1. The
parameter $V_0$ denotes the activation energy for ionic diffusion. The new profile $u(x,t)$ is updated from
the migration rates, and from (\ref{resistance}) 
we get the new total (two point) $R(t)$ as the discrete $x$-integral of the local cell's resistance $r(x,t)$. 
Here, for simplicity, we focus on a single active SB-bulk interface, while the more general situation with two barriers may be 
analyzed following a similar line of argument \cite{prb}. 
The applied external electric stress that we adopt is a constant current, in both, simulations and experiments (see below). 

As described in Ref.\onlinecite{prb}, the initial vacancy concentration profile is assumed to be constant $[u(x)]=[u^0]$. 
The ``forming'' or initialization of the memory is done by first applying a few 
current loops of alternative polarity, $\pm I^0$, until the migration of vacancies evolves towards a {\em limit cycle},
with a well defined profile $u_0(x)$.
After this, the system begins to repetitively switch between two values: $R_{HI}$ and $R_{LO}$.
In the first, most of the vacancies reside within the high-resistance region SB, and in the second they accumulate 
vacancies in the more conductive bulk. 
The $R_{HI}$ state with the vacancies piled up in the first cell, at $x$=1, defines the initial state for 
the shock wave propagation (see Fig.\ref{fig:shock wave dynamics}).

{\em Shock wave formation: the "propagation phase".} 
We apply an external field with polarity pointing from the SB to the bulk and observe the evolution of the vacancy profile as a function of
the (simulation) time. As can be observed in Fig.\ref{fig:shock wave dynamics} there is a rapid evolution of the profile 
into a shock wave form with a sharply defined front. 
We also notice that the total resistance remains approximately constant during an initial phase,
and suddenly starts to decrease after the front hits the internal SB-bulk interface at $x_{int}$ (inset of 
top panel of Fig.\ref{fig:shock wave dynamics}).  
We shall analyze these key features in the following.

First, we focus on the propagation of the shock wave front position $x_s(t)$, as shown in Fig.\ref{fig:shock wave dynamics} (bottom-left panel) 
for different values of the electronic current $I$.  We observe that the characteristic time $\tau_1$  for the shock wave to travel through the
Schottky barrier and reach the SB-bulk point $x_{int}$ decreases exponentially with the magnitude of $I$. 
To obtain analytical insight for this behaviour, we recall that the velocity of the shock wave front $dx_{s}/dt$ is very generally given by
the Rankine\textendash{}Hugoniot conditions \cite{debnath2011nonlinear,landau1987fluid}, which express it 
as the ratio of the spatial discontinuity of the (vacancy) drift current, and the spatial discontinuity of the density 
profile across the shock viz. $dx_{s}/dt=\Delta j/\Delta u\left|_{x_{s}}\right.$.
Within the VEOVM model \cite{prb}, we  obtain the following {\em nonlinear rate equation} 
 (see Supplementary Materials for details), which describes the dynamics of the shock wave front:
\begin{eqnarray}
\frac{dx_{s}}{dt} & = & \frac{2Du_{-}\sinh\left(IA_{S}u_{-}\right)-2Du_{+}\sinh\left(IA_{S}u_{+}\right)}{\Delta u}, 
\label{eq:shock wave velocity}
\end{eqnarray}
where $D$ is a prefactor related to the activation energy for vacancy migration (Arrhenius factor) (see Eq.\ref{burgers} 
and Ref.\onlinecite{prb}), 
and $u_{-/+}$ are the density of vacancies at the two sides of the shock wave front (see Fig.\ref{fig:shock wave dynamics}). 
The density $u_{-}$ depends on the shock wave front position via: $u_{-}= Q/x_{s} + u_{+} $, where $Q$ is the 
total number of vacancies carried by the shockwave, which remains a constant parameter
through the propagation phase ($t <\tau_1$) and $u_{+}$ is a constant 
background density which can be written as $u_{+}=Q_{B}/x_{int}$, $Q_{B}$ standing for the total number of background vacancies.  

Our description of the propagation phase is fully consistent with our numerical
simulations. As shown in the inset of Fig.\ref{fig:shock wave dynamics} (top panel), the resistance remains 
essentially constant until the wave front reaches the SB-bulk interface after a (current dependent) time $\tau_1$, 
and then begins to drop. Moreover, we also achieved a good fit to the shock front velocity
by using Eq.\ref{eq:shock wave velocity}, 
as is shown in Fig.\ref{fig:shock wave dynamics} (see Sup. Mat. for details). 

\begin{figure}[h]
 \includegraphics[width=0.49\textwidth]{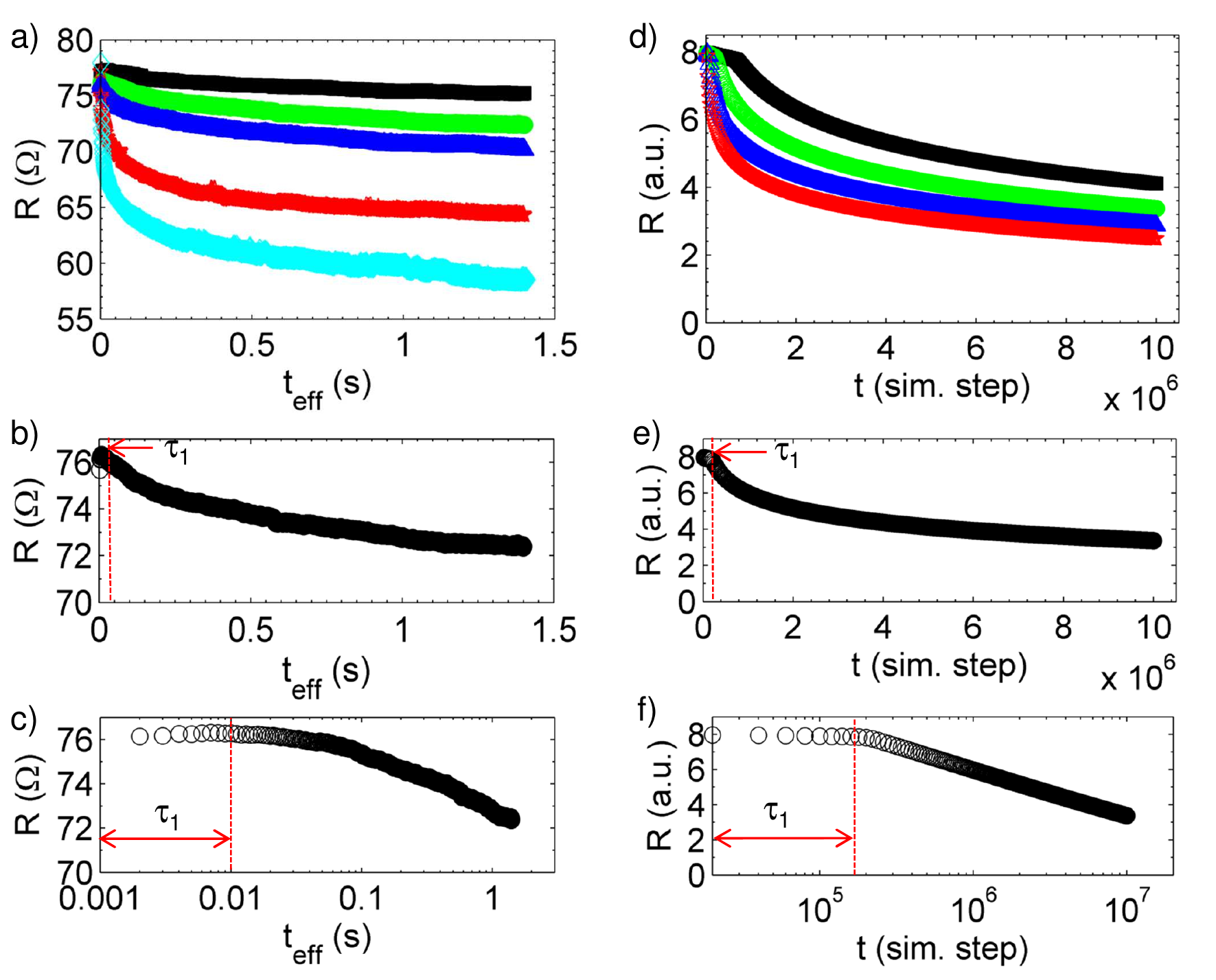}
\caption{ Time dependence of the resistive change $R(t)$ for various external current intensities.
Top left: experimental data measured on an Ag/LPCMO/Ag memristor with $I$ = 37.5 mA (black), 40 mA (green), 50 mA (blue), 
80 mA (red) and 100 mA (cyan).  $t_{eff}$ is the effective time duration of the applied currents (see Supplementary Material). 
The initial state was reset by applying an intense negative polarity current of 350 mA. Note that the initial value of the
resistance, 78 $\Omega$ is recovered within $\pm 1 \Omega$.
Top right: model simulations with applied current: $I[a.u.]$ = 58.5 (black), 71.5 (green), 84.5 (blue) and 97.5 (red).
Middle panels: experimental data for $I$ = 40 mA (left) and simulations for $I[a.u.]$ = 71.5 (right).
$\tau_1$ is defined as the time interval from the beginning of the applied pulse until the resistance starts to drop.
Bottom panels: idem in a semi-log plot.}
\label{fig.R(t)}
\end{figure}

{\em Resistance switching: the "leakage phase".}
After the shock front reaches the boundary point $x_{int}$ the resistance begin to drop. To understand this 
behaviour, we note that the total resistance of the Schottky barrier is given by the {\em  total number 
of vacancies} within the barrier region viz. from (\ref{resistance}) $\ensuremath{R_{SB}=\int_{SB}dxA_{S}u(x)}$. As a result, the 
resistance drop per unit time is approximately given by the ionic vacancy-current passing through the SB-bulk 
interface at $x_{int}$,
\begin{eqnarray}
dR(t)/dt = -A_{S}j(x=x_{int})
\end{eqnarray}
since $R_S \gg R_B$ as $A_S \gg A_B$.
Notice that during the propagation phase the ionic current through the interface $x_{int}$ is negligibly small. 
This is because the initial vacancy concentration there, and hence the local field, are also negligibly small. 
However, when the shock wave front eventually reaches 
the end of the SB region, after travelling for a time $\tau_{1}$, we do expect a sudden resistance 
drop as a large number of ionic vacancies begin to leak out into the bulk region.

We shall now focus on the detailed description of  the resistive drop. In Fig.\ref{fig.R(t)} we show 
the systematic dependence of $R(t)$ as a function of the applied external (electronic) current.
Along with the simulations of the VEOVM, we also present our experimental results measured on 
a manganite-based (La$_{0.325}$Pr$_{0.300}$Ca$_{0.375}$MnO$_3$) memristive device.
Experimental details are provided in the Supplemental Material. 
The set of curves were obtained for applied current intensities just above the threshold for the onset of the resistance switch. 
The goal was not to demonstrate the fast switching speed of the device, but rather on the contrary, achieve
relatively slow switching speeds in order to access the different time scales. We observe that in both, simulation 
and experiments, the resistance change rapidly becomes larger and faster with the increase of the applied electric stress
intensity.
We also observe an overall good qualitative agreement between experiments and model simulations. This 
is also highlighted by the semi-log plots, which clearly display the two-stage process 
involved in the resistive switch, before and after the impact time $\tau_1$.

Remarkably, within shock wave scenario, we may also obtain explicit expressions that quantify 
the resistance change during the leakage phase. Our analysis may be simplified by first noting, 
from general considerations of shock waves, that their shape at long times becomes "flat", i.e. 
the gradient of the local density rapidly decreases ($\partial_{x}u\rightarrow0$) at all points 
that were overtaken by the shock wave front\cite{debnath2011nonlinear,courant1962methods}.  
Indeed, our data is fully consistent with this 
observation, as the  vacancy density profile within the SB remains approximately "flat" (ie spatially constant $u(x,t) = u_S(t)$) 
at all times after the shock front reaches the interface (see Fig.\ref{fig:shock wave dynamics}). 
Then, within the VEOVM the SB resistance is simply proportional to the total vacancy concentration within the barrier and we
have,  $R(t) \approx R_S(t) = A_{S}x_{int}u_{S}(t)$. Since the electronic current $I$ is held fixed, 
the vacancy (i.e. ionic) current through the interface depends only on the vacancy concentration $u_S$ 
(cf Eq.\ref{model}). Thus, within the VEOVM we obtain a {\em nonlinear rate equation}, 
describing the resistance drop during the "leakage phase":
\begin{eqnarray}
\frac{dR}{dt}  = -\frac{2DR}{x_{int}}\sinh\left(\frac{IR}{x_{int}}\right).\label{eq:R(t)}
\end{eqnarray}
Similarly as we showed before for the propagation phase, this equation may be validated by
a quantitative fit to the simulation results (see Sup. Mat.). 
Note that due to the strong nonlinear form of this rate equation, the $R(t)$ response is significantly different 
from the simple exponential decay expected in the familiar linear case (e.g. in standard RC circuits). 
Therefore, within the short time scale associated with the initial fast drop of resistance
and where the RS is significant ($IR/x_{int}\gg1$), 
the present type of nonlinear system is dominated by the activated process and
the approximation $\sinh\left(IR/x_{int}\right)\approx\frac{1}{2}\exp\left(IR/x_{int}\right)$ is valid.
This enables the approximate solution of the Eq.\ref{eq:R(t)}. 
\begin{eqnarray}
R  =  R_{HI}-\frac{x_{int}}{I}\ln\left(1+\frac{t}{\tau_2\left(I\right)}\right),\label{eq:approximate solution}
\end{eqnarray}
where the time is measured from the ``impact'' time $\tau_{1}$ and
$\tau_2\left(I\right)= \frac{x_{int}^2}{DIR_{HI}} \exp\left(-IR_{HI}/x_{int}\right)$ (see Sup. Mat.) 
is the current-dependent characteristic time for the resistance drop. 
%

{\em Resistivity scaling.}
An interesting consequence of Eq.\ref{eq:approximate solution} is that it suggests
the scaling behaviour of the curves $R(t)$.  
In fact, one may define the normalized resistance drop 
$\delta R\left(t^{*}\right)=R-R\left(\tau_{2}\right)/\left(R_{HI}-R\left(\tau_{2}\right)\right)$ 
and see from Eq.\ref{eq:approximate solution} that obeys it the scaling form:
\begin{eqnarray}
\delta R\left(t^{*}\right)=1-\ln\left(1+t^{*}/\tau_{2}\right)/\ln\left(2\right), \label{eq:scaling}
\end{eqnarray}
 
\begin{figure}[h!]
\includegraphics[width=0.5\textwidth]{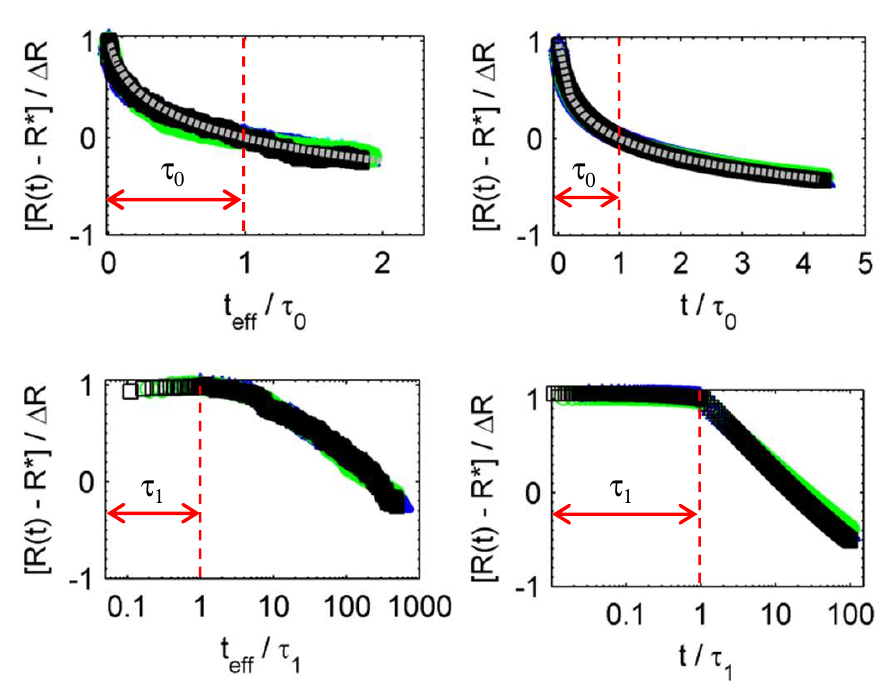}
\caption{ Scaled curves of the $R(t)$ data sets of Fig.\ref{fig.R(t)}.
The left panels show the collapsed experimental data and the right ones the numerical simulations.
The time $\tau_0$
is an auxiliary scaling variable, which
is proportional to the characteristic time $\tau_2$ (see Sup. Mat. for details on the
scaling procedure).
The scaled data were fitted (white dotted line) with a generalized version of Eq.\ref{eq:scaling}
(see Sup. Mat.).
The lower panels show the same data sets scaled with the
shock wave impact time $\tau_1$ (the experimental curves show only three
data sets for the lower current values. At higher currents our electronics could not resolve $\tau_1$). 
To achieve the scaling of the lower panels, we assumed for each plot the normalization
value of $\Delta R$ determined from the previous scaling (top panels).
}
\label{scaling}
\end{figure}

In Fig.\ref{scaling} we demonstrate that this striking feature is indeed present 
in both, our experiments and simulations data.
In the upper panels of the figure we show the excellent scaling that is achieved, where all the 
experimental and the simulation curves $R(t)$ from Fig.\ref{fig.R(t)} were respectively collapsed 
onto a single one.
Moreover, the collapsed data can also be fitted with a slightly more general form of 
Eq.\ref{eq:scaling}, that we discuss in the Supplemental Material.
Remarkably, in the lower panels of Fig.\ref{scaling} we show that a collapse of the data $R(t)$ can also be obtained
using the impact time $\tau_1$ as the scaling variable. This is significant, because it shows that a single scaling behavior
may include the two phases of the resistive switching process, namely before and after $\tau_1$.
We should mention that the scaling scenario was derived with the assumption of an ohmic behaviour 
in the $I$-$V$ characteristics. While this may not be the case in general \cite{FGM}, within the present set
of experiments, which are performed near the current threshold of RS, our results indicate that this is a reasonable 
assumption or at least a valid approximation.

To conclude, from quite general considerations of migration of ionic defects under strong electric fields
in solids, we have argued that the dynamics of the spatial profile of defect concentration should be governed
by a Burgers'-type nonlinear equation and develop shock waves.
We demonstrated that this scenario is indeed realized within a concrete realization, namely
a ionic migration model that was previously applied to describe resistive switching phenomena in manganite 
based memristive devices. In those systems, a key role is played by the migration of oxygen vacancies,
which are the ionic defects relevant to the electronic transport properties.
We thus predicted a two-stage process for the resistive switch phenomenon. An initial one, where the
oxygen-vacancy concentration profile develops a shock wave that propagates throughout a highly 
resistive (Schottky barrier) region near the electrode. During this phase the resistance essentially
does not change. 
This is followed by a second phase, where the shock wave emerges from the high resistive region and 
the ionic defects leak into the conductive bulk.
Our scenario was further validated by novel experimental data on a manganite based memristor device.
A remarkable results of our study is that both, the numerical simulations and the experimental curves,
obeyed a scaling behaviour, providing decisive support to our theory.
The present work provides novel insights on the physical mechanism behind the commutation speed on 
novel non-volatile electronic memories,  unveiling an unexpected connection between a phenomenon
of technological relevance and a classic theme of nonlinear dynamical systems.

This work was partially supported by public grants from the French National Research 
Agency (ANR), project LACUNES No ANR-13-BS04-0006-01, the NSF 
DMR-1005751and DMR-1410132, the University of Buenos
Aires (UBACyT 2013-2016) and the Conicet PIP-2013-MeMO.

\bibliographystyle{apsrev}

\end{document}


\title{Supplemental Material for Shock-waves and commutation speed of memristors}

\author{Shao Tang}

\affiliation{Department of Physics and National High Magnetic Field Laboratory,
Florida State University, Tallahassee, FL 32306, USA}

\author{Federico Tesler}

\affiliation{Departamento de F\'{i}sica - IFIBA, FCEN, Universidad de Buenos Aires, Ciudad
Universitaria Pabell\'on I, (1428) Buenos Aires, Argentina}

\author{Fernando Gomez Marlasca}

\affiliation{GIA-CAC - CNEA, Av. Gral Paz 1499 (1650) San Mart\'{i}n, Pcia Buenos Aires, Argentina.}

\author{Pablo Levy}

\affiliation{GIA-CAC - CNEA, Av. Gral Paz 1499 (1650) San Mart\'{i}n, Pcia Buenos Aires, Argentina.}

\author{V. Dobrosavljevi\'{c}}

\affiliation{Department of Physics and National High Magnetic Field Laboratory,
Florida State University, Tallahassee, FL 32306, USA}

\author{Marcelo Rozenberg}

\affiliation{Laboratoire de Physique des Solides, CNRS-UMR8502, Universit\'e de
Paris-Sud, Orsay 91405, France}


\maketitle

\section{Method of characteristics}

In this section, we demonstrate that shock wave are formed in general
as a consequence of the nonlinearity. We analyze the problem under
generic circumstances via the \textit{method of characteristics}.
We then provide an explicit example which is qualitatively different
from the situations discussed in main text. 

\begin{figure}[H]
\includegraphics[width=3.03in,height=3.03in,keepaspectratio]{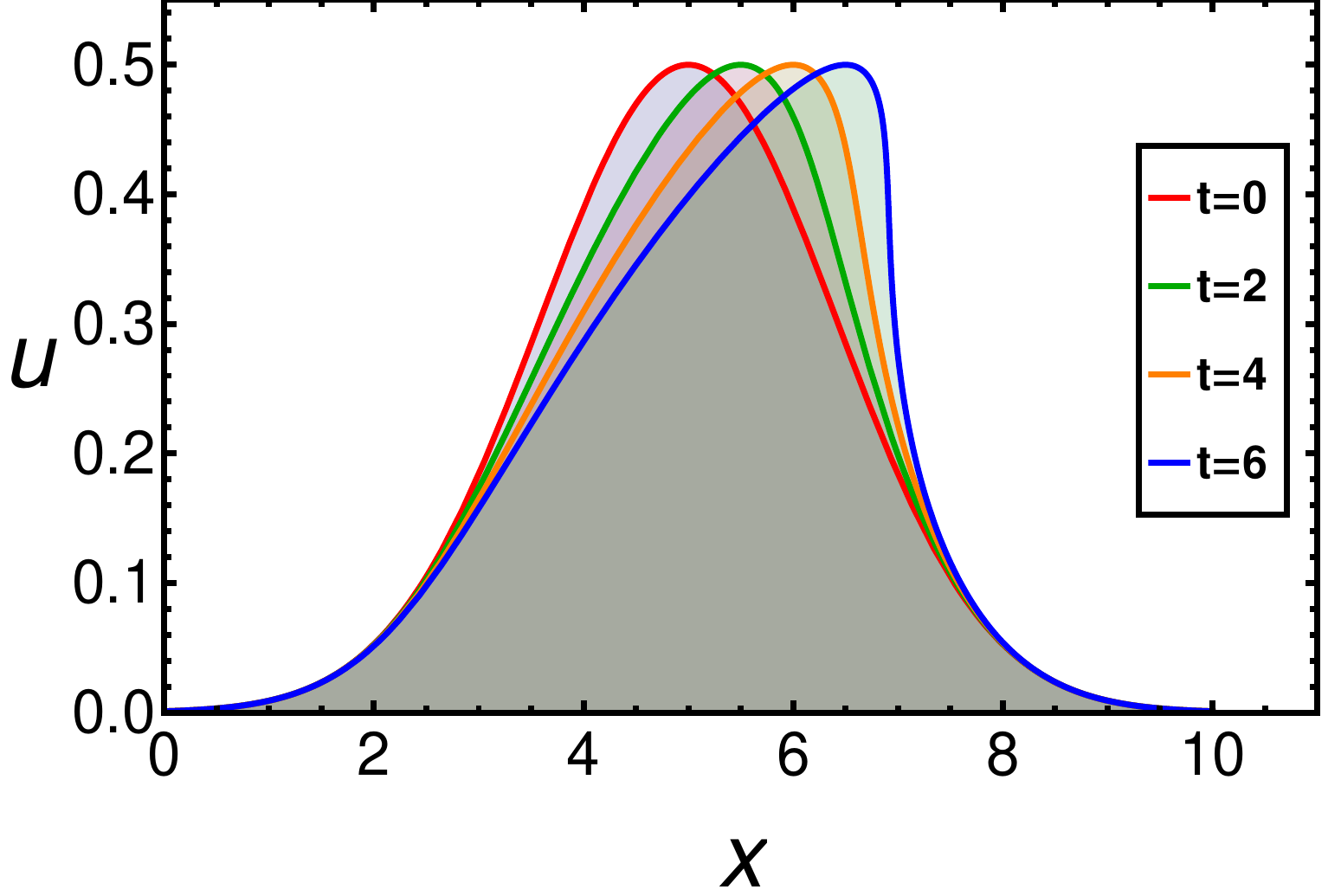}\includegraphics[clip,width=3in,height=3in,keepaspectratio]{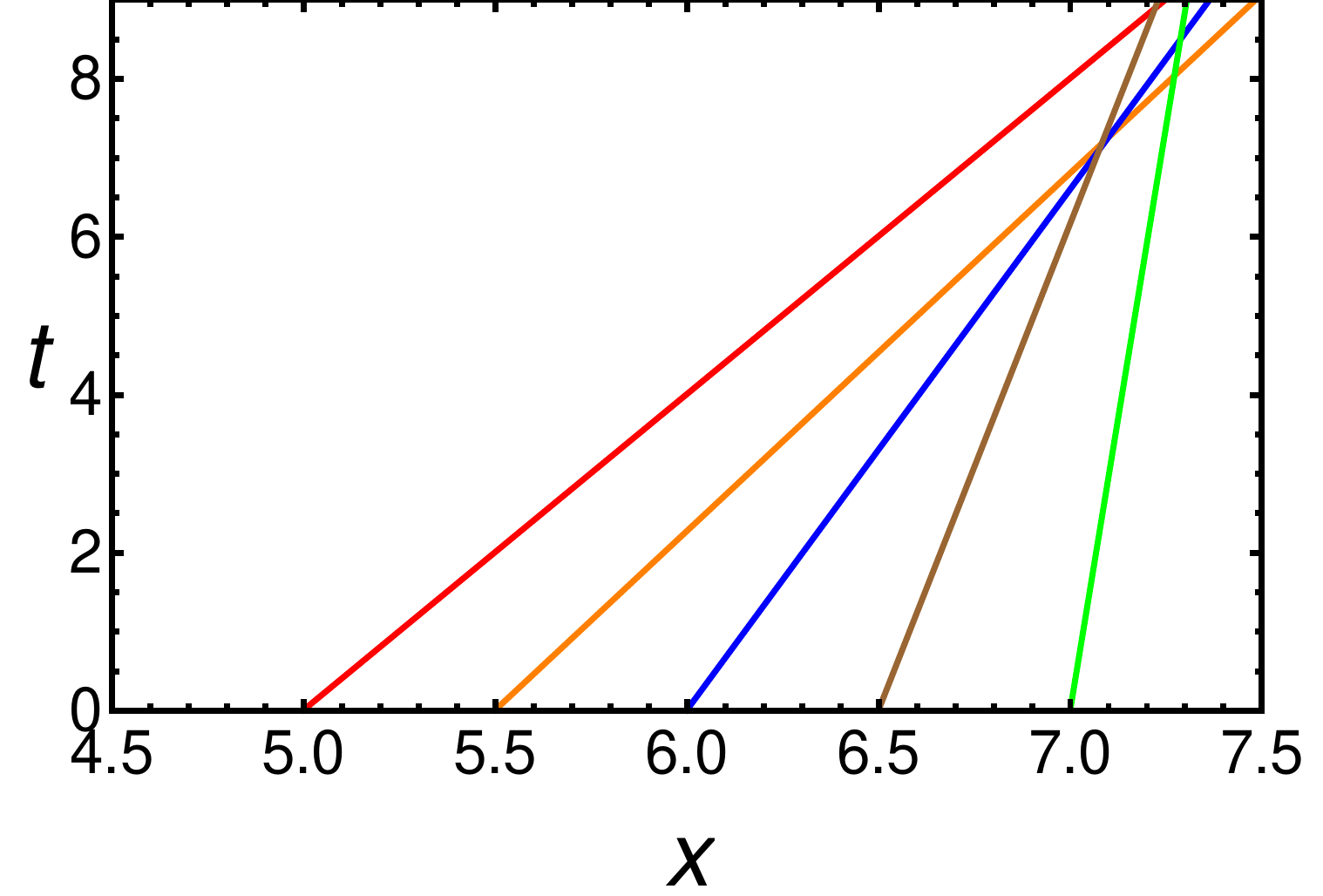}

\footnotesize \caption{\label{fig:Systematic-illustration-for}Left panel: The motion of
the wave and the shock with steep wave front forms at $t=6$ approximately.
Right panel: Corresponding characteristics start at $x_{0}=5,5.5,6,6.5,7$
separately and crosses happen around $t=6$, indicating the onset
of shock wave.}

\end{figure}

Considering the vacancy migration in high resistive barriers where
we can neglect the normal diffusion term $D\partial_{xx}u$, we would
then obtain that the first order, quasi-linear partial differential equation
has the form $a(u,t,x)\partial_{t}u+c(u,t,x)\partial_{x}u=b(u,t,x)$
in general and the $u$ dependence in $c(u)$ is responsible for the
onset of the shock wave. The characteristics\cite{debnath2011nonlinear,courant1962methods}
are defined as (using $s$ as a parameter along these curves):
\begin{equation}
\begin{cases}
\frac{dt}{ds}=a(u,t,x)\\
\frac{dx}{ds}=c(u,t,x)\\
\frac{du}{ds}=b(u,t,x)
\end{cases}\label{eq:characteristics}
\end{equation}
We have the following equation as the ion\textquoteright{}s motion
in barrier is concerned:

\begin{equation}
\partial_{t}u+c(u,t)\partial_{x}u=0\label{Burger's type eq}
\end{equation}
We can immediately write down the characteristic equations which are
simply $du/dt=0$ and $dx/dt=c(u,t)$ with $dt/ds=1$. It's easy to
see that $u$ is a constant along the characteristics since $du/dt=dx/dt\partial_{x}u+\partial_{t}u=0$.
Starting from the initial data, different lines might intersect with
each other, indicating different values of $u(t=0,x)$ which propagates
from initial position meet each other hence the $u$ at this point
on $t-x$ plane is multi-valued. The intersection which happens first
chronologically determines the onset of the shock wave as shown in
Fig.S\ref{fig:Systematic-illustration-for}.

We would elaborate in a more physical way as follows with the nonlinear
drift term which satisfies the condition $\partial_{u}c(u,t)>0$ .
Considering (\ref{Burger's type eq}) as a simple wave equation, it's
obvious that a travelling wave with speed locally proportional to $c$ 
is the solution. (Compared to a more basic situation where $c(u,t)$
is independent of $u$, then we obtain a travelling  wave with the same
speed for every point.) Consequently, the characteristics defined
above records the path as every point with a definite $u$ moves in
$t-x$ plane. Therefore, at any given time, spatially different points
move with the different speed proportional to $c(u,t)$ and the ``crest''
with the largest value of $u$ move fastest which leads to a ``kink''
type shock. It's also straightforward to check that if $\partial_{u}c(u,t)\leq0$,
there would be no shock with steep wave front formed. In the case
where $c(u,t)$ is independent of $u$, the characteristics are curves
with completely the same shape, paralleling to each other and hence
would not have any intersections. Considering the polarity with drift
force pointing from left barrier to conductive bulk, as long as the
initial data $u(t=0,x)$ is \textit{not everywhere non-decreasing}
(there exists two points $x_{1}<x_{2}$ with initial data satisfies
$u(t=0,x_{1})>u(t=0,x_{2})$), the shock wave would be established
after some time since $\partial_{u}c(u,t)>0$. 

Therefore, upon reversing the polarity, most of the vacancies accumulates
in the left end of the barrier with $u(t=0,x)$ to be a strong decreasing
function, experiencing the significant drift force pointing to the
conductive bulk which guarantees the quick onset of the shock wave
front (kink) with relative large amplitude.

As an illustration, we provide an example with $c(u,t)=u^{2}$ and
the Gaussian distributed initial data as $0.5\exp\left[\left(x-5\right)^{2}/4\right]$.
The time dependent wave profile is solved numerically with fixed boundary
condition with characteristics computed according to Eq.\ref{eq:characteristics}.

\section{Dynamics of shock wave }

In this section, we provide a simple derivation of the Rankine\textendash{}Hugoniot
conditions\cite{debnath2011nonlinear,landau1987fluid} which determines
the equation of motion of the shock wave front. Notice that an important
feature of the shock is the spatial discontinuity of $u$ and also
$j$. Considering a shock wave front propagating between the interval
$\left[0,d\right]$ in 1D, we then define $u_{+}\equiv u(t,x_{s}(t)+\epsilon)$,
$u_{-}\equiv u(t,x_{s}(t)-\epsilon)$, $j_{+}\equiv j(t,x_{s}(t)+\epsilon)$
and $j_{-}\equiv j(t,x_{s}(t)-\epsilon)$ correspondingly with $x_{s}(t)$
to be the coordinate of the shock wave front and $\epsilon\rightarrow0^{+}$
is a infinitesimal positive quantity. According to the continuity
equation $\partial_{t}u+\partial_{x}j=0$, we have:

\begin{equation}
\frac{d}{dt}\left(\int_{0}^{x_{s}(t)}udx+\int_{x_{s}(t)}^{d}udx\right)=j(0)-j(d)
\end{equation}

\begin{eqnarray}
\frac{dx_{s}}{dt}u_{-}-\frac{dx_{s}}{dt}u_{+}+\int_{0}^{x_{s}(t)}\partial_{t}udx+\int_{x_{s}(t)}^{d}\partial_{t}udx & = & j(0)-j(d)
\end{eqnarray}
Applying the continuity equation again, we obtain the shock wave velocity
as follows:

\begin{eqnarray}
v_{s}=\frac{dx_{s}}{dt}  =\frac{j_{+}-j_{-}}{u_{+}-u_{-}}=  \frac{\Delta j}{\Delta u}\left|_{x_{s}}\right.\label{eq:shock wave velocity general}
\end{eqnarray}
Therefore, we obtain the equation of motion regarding the shock wave
front which is valid for a general form of $j(u,t,x)$. 

\section{VEOVM model }

As it was mentioned in the main text, to simulate the vacancies dynamics we adopted \emph{the voltage enhanced oxygen vacancy migration model} (VEOVM), which is a well validated model for the RS effect. Taking a capacitor-like device, this model considers a 1 dimensional conductive channel connecting the two contacts, along which oxygen vacancies can migrate through (see Fig.S2). This channel is divided in small cells corresponding to physical nano-domains, and at the same time the whole device is divided in two regions: the active interfacial region close to the metal contact (i.e. high resistive Schottky Barrier) and the high conductive bulk region. A diagram of the model with a single active contact, as used in the simulations, is presented at the top panel on Fig.\ref{fig:Fig-Sim}. The resistance of each cell in the channel $r(x)$ is proportional to the local density of vacancies: $r(x)=u(x)A_{\alpha}$, where $u(x)$ is the density of vacancies on the cell located at $x$ and $A_{\alpha}$ is a proportionality constant whose magnitude depends on the region of the device: $\alpha=S,B$, where $S$ stands for Schottky barrier and $B$ for bulk, 
and where $A_{s}\gg A_{B}$.
Migration is assumed to occur only between neighbours cells, and in each step of the simulation the probability for vacancy migration from a cell at $x$ to its neighbour at $x+\Delta x$ is computed according to:

\begin{equation}
P(x, x+\Delta x)=u(x)(1-u(x+\Delta x))\exp(\Delta V(x)-V_0),
\label{Prob}
\end{equation}

where $\Delta V(x)$ is the drop of voltage (per unit length) at $x$ and $V_0$ is an activation constant for vacancy migration. In each step of the simulation the migration from cell to cell is computed and the new resistance of each cell is calculated. The total resistance of the device is calculated simply as the addition of all the cells in the channel. For the values of used in the simulations, the middle term $(1-u(x+\Delta x))$ can and will be neglected in the theoretical analysis explained in the following sections.
As mentioned in the main text, this model corresponds to the case of granular materials with activated transport process. To put it in the context of the $generalized-Burgers' equation$ description made in main text, it is easy to see that the drift current originated from this model satisfies the general form $j_{drift}\sim \sinh\left(E\right)$. The net current of vacancies generated by the action of an external electric field $E(x)=\partial V(x)/\partial x$, between a cell at $x$ and its neighbour cell at $x+\Delta x$  can be written simply as Fick's-law for the migration probability: $j_{drift}= \partial P/\partial x \approx [P(x,x+\Delta x) -P(x+\Delta x,x))]/\Delta x$ , where in the last term the discrete character of the model has been introduced. In particular for our model we take $\Delta x=1$. Using the definition of $P$ from the model (neglecting its middle term), we get from here that $j_{drift}=P(x, x+\Delta x) - P( x+\Delta x,x) = 2Du\sinh\left(\Delta V(x)\right) $, where $D$ stands for the Arrhenius factor $\exp(-V_0)$. If the external electric stress is a controlled current $I$, then $\Delta V(x)=Ir(x)=IA_{\alpha}u(x)$. 

\section{Simulations details}

\begin{figure}[h]
  \includegraphics[width=0.5\textwidth]{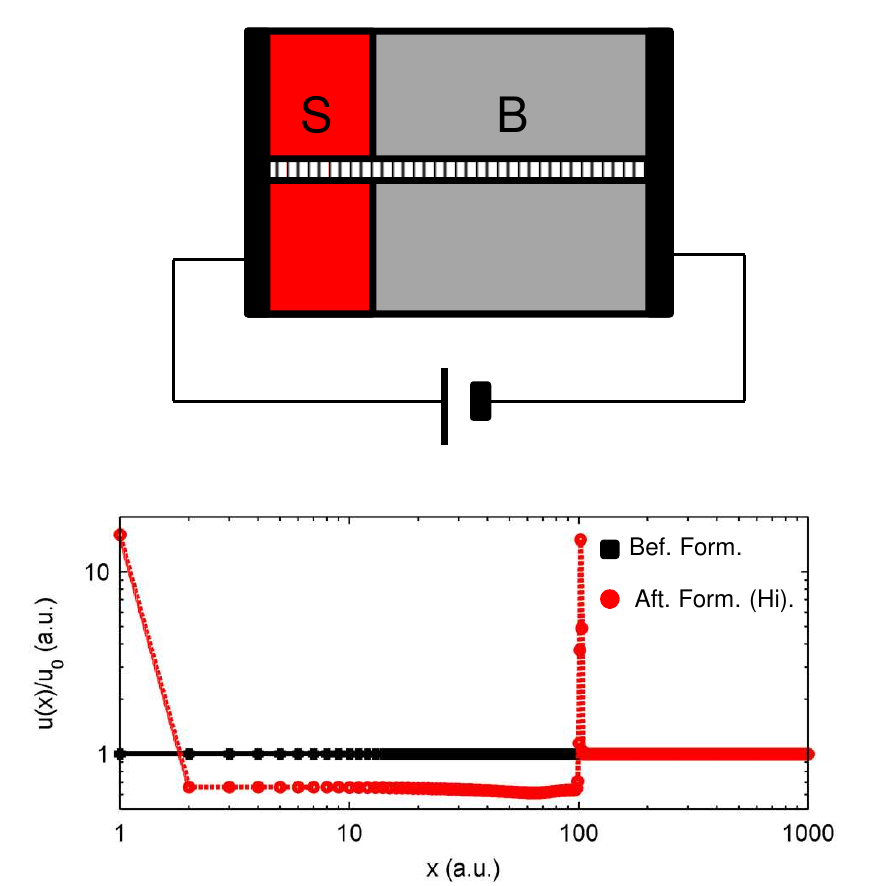}
  \caption{ Top panel: Schematic diagram of the VEOVM model with a single active contact. The two regions S and B correspond to the high resistance interface (Schottky Barrier) and the more conductive central bulk, respectively. The small cells within the channel indicate the domains. Bottom panel: vacancy distribution along the conductive channel before (black squares) and after (red circles) the forming process. The distribution shown after the forming process correspond to a Hi resistance state. Starting from a uniform distribution, during the forming process a characteristic distribution is obtained where vacancies initially at the interface migrates to the low-field bulk region generating a pile-up of vacancies close to the limit between the two regions. In the Hi state, part of these vacancies migrates back to the interfacial region and are accumulated in the vicinity of to the metal contact.
   }
  
  \label{fig:Fig-Sim}
\end{figure}

In order to analyze the existence of a shock wave scenario, the Hi to Lo process was simulated for different external currents. Previously, an initial forming process was performed at which current of different polarities was applied and well defined Hi and Lo states were obtained. In the bottom panel on Fig.\ref{fig:Fig-Sim} the vacancy profile is shown before and after the forming process. The device starts with a uniform distribution along the channel, and by the end of the forming process the profile shows a characteristic distribution in which vacancies that were originally in the interfacial region migrates to the low-field bulk region where they are accumulated. The figure shows the formed Hi resistance state where it can be seen a second accumulation of vacancies next to the left metal contact in the interfacial region. This second accumulation of vacancies conforms the initial distribution for the Hi to Lo process and it will evolve to form the shock wave during switching as shown in the main text.
To improve the stability of the simulations the currents were applied using a rise time (from 0 to their actual values) always negligible compared with the characteristic times of the process $\tau_{1}$ and $\tau_{2}$ (with rise times in the order of a few thousands of steps).
The simulated device contained a total of 1000 cells with a 100 cells long interfacial region (left interface). For the initial distribution a uniform density of $u_0=1\times10^4$ per cell was used. The value of the activation constant $V_0$ was set to 16, and for the resistivity constants we used $A_S=1000$ and $A_B=1$.

\section{fitting for VEOVM model }

In this section, we explain the details of the comparison between simulation
and theory and the procedures used for fitting. Consider the drift current within
left interfacial region in VEOVM model:

\begin{eqnarray}
j(u,t,x) = P(x, x+\Delta x) - P( x+\Delta x,x) = 2Du\sinh\left(IA_{S}u\right)
\end{eqnarray}

where the term $1-u(x+\Delta x)$ in the probability  has been neglected as explained before \footnote{The typical value of $u(x)$ in simulation varies between $10^{-4}$ and $10^{-3}$.}.
From Eq.~(\ref{eq:shock wave velocity}) we have, as shown in the main text:

\begin{eqnarray}
\frac{dx_{s}}{dt} & = & \frac{2Du_{-}\sinh\left(IA_{S}u_{-}\right)-2Du_{+}\sinh\left(IA_{S}u_{+}\right)}{\Delta u},
\end{eqnarray}

where $x_s$ is the shock wave front position, $u_{-}$ is the density of vacancies inside the shockwave (assumed uniform), $u_+$ is the density outside the shockwave (background vacancies), $I$ is the applied current, and $D$ is the Arrhenius factor $\exp(-V_0)$.

Performing a redefinition of parameters (for practical reasons only) we can rewrite the last equation as:

\begin{eqnarray}
\frac{dx}{dt} =  \frac{2D}{x_{int}}\left[\left(1+\frac{\alpha}{\beta}x\right)\sinh\left(\frac{I}{\beta}+\frac{I}{\alpha x}\right)-\frac{\alpha}{\beta}x\sinh\left(\frac{I}{\beta}\right)\right],
\label{eq:shock wave velocity}
\end{eqnarray}

where we used that $u_{-}=u_{+}+\Delta u$ , $Q=\Delta u x_{s}$ is the total number of vacancies carried
by shock wave, $Q_{B}\equiv u_{+}x_{int}$ is total number of background of vacancies in the left region (being $x_{int}$ the length of the interfacial region),  $x\equiv x_{s}/x_{int}$ is the normalized coordinate, and where we defined the new parameters $\alpha \equiv x_{int}/A_{S}Q$ and $\beta \equiv x_{int}/A_{S}Q_{B}$.
From the previous definitions we can write the high resistance value as $R_{HI}=A_{S}\left(Q_{B}+Q\right)$
where $R_{HI}$ is a constant determined by the vacancy concentration and independent of $I$
 \footnote{There is in fact a small migration of background vacancies into the bulk during the period of shockwave propagation which causes a small variation of the resistance during this phase.}.

We can solve then this equation numerically with the material dependent parameters $x_{int}$, $\alpha$,$\beta$ and $D$. 
Considering the initial rise time for the current and the non-flatness of the shock wave, an accurate test of the $x_s(t)$ prediction can be performed for $x>x_{0}\approx0.4$ using the integral-form equation: 

\begin{eqnarray}
t-t_{0} & = & \intop_{x_{0}}^{x}\frac{\left(x_{int}/D\right)dy}{2\left(1+\frac{\alpha}{\beta}y\right)\sinh\left(\frac{I}{\beta}+\frac{I}{\alpha y}\right)-\frac{2\alpha}{\beta}y\sinh\left(\frac{I}{\beta}\right)}\label{eq:t-x simulation}
\end{eqnarray}

which is used for the fit of the simulation data for $t <\tau_1$ in the lower left panel of Fig.1 in the main text.

On the other hand, we have also the rate equation for the switching (i.e. "leakage") phase 
(cf. Eq.4 from main text):

\begin{eqnarray}
\frac{dR}{dt} & = & -\frac{2DR}{x_{int}}\sinh\left(\frac{IR}{x_{int}}\right)\label{eq:R(t)}
\end{eqnarray}

The numerical solution of this equation is used to fit the simulation data in the resistance switching phase
for $\tau_1 < t $, shown in Fig.S3.

The values of the parameters that enter the equations (\ref{eq:t-x simulation}) and (\ref{eq:R(t)}) were extracted directly 
from the simulation results (Fig.1 of main text). We find $Q_{B}=63.4u_0$, $Q=15.3u_0$, $A_S=1000$, $R_{HI}=7.9a.u.$ 
and $D=1.12\times 10^{-7}$. The very good fits of the simulation data shown in Fig.1 and Fig.S3 were achieved by slightly 
relaxing  the value of the single parameter $x_{int}$=100 to the values 94.3 (Fig.1) in the propagation phase
and 82.2 (Fig.S3) in the leakage phase.

The small discrepancy between the relaxed parameters and its actual value mainly comes from the non-flatness of the density 
profile both during the propagation of the shock wave and during the leaking phase, as well as from the fact that a small part 
of background vacancies leaks into the bulk during the propagating phase.

\begin{figure}[h]

	\includegraphics[width=0.48\columnwidth]{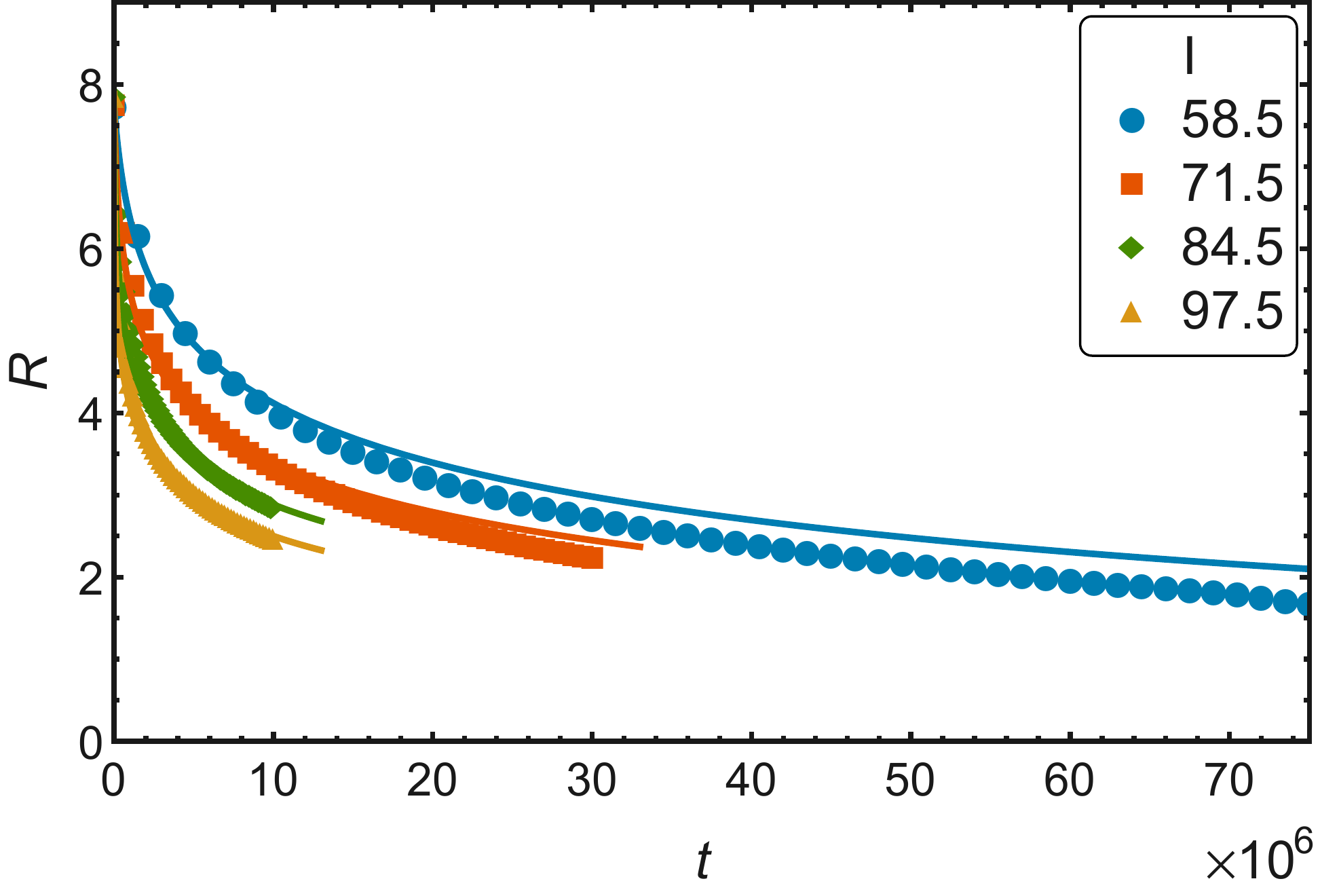}

\caption{Evolution of the resistance during the switching phase for different applied currents 
according to simulations (dots) and theory (lines) from Eq.~(\ref{eq:R(t)}). 
The currents shown are $I=58.5a.u., 71.5a.u.,84.5a.u.$ and $97.5a.u.$}
\label{fig:R-fit}
 \end{figure}

\section{Scaling}

In this section, we show that the scaling behaviour
is a direct consequence of the strong non-linearity of the drift current
$j(u,t,x)$, which depends on local electric field exponentially. 
In the time range where significant resistive switching occurs, we have ($IR/x_{int}\gg1$)
and we may approximate $\sinh\left(IR/x_{int}\right)\approx\frac{1}{2}\exp\left(IR/x_{int}\right)$.
From Eq.(\ref{eq:R(t)}), using a normalized resistance $\widetilde{R}=R/R_{HI}$,
we have:

\begin{eqnarray}
\frac{d\widetilde{R}}{dt} & = & -\frac{D}{x_{int}}\exp\left[\left(\frac{IR_{HI}\widetilde{R}}{x_{int}}\right)\left(1+\lambda\right)\right],
\end{eqnarray}
where $\lambda=\frac{x_{int}}{R_{HI}}\ln\widetilde{R}/\left(I\widetilde{R}\right)$
is a small parameter as long as $\widetilde{R}$ close is to $1$ (i.e. at the beginning of the resistive change). 
For simplicity we consider the leading order as we set $\lambda=0$:

\begin{eqnarray}
\widetilde{R} & = & 1-\frac{x_{int}}{IR_{HI}}\ln\left(1+\frac{t^{*}}{\tau_2}\right),\label{eq:approximate solution}
\end{eqnarray}

where $t^{*}= t-\tau_1$ is the time measured from the impact time as explained in main text, and  $\tau_2$ is a characteristic time for the resistance switch and is dominated by an exponential dependence
of the applied current as follows: 

\begin{equation}
\tau_2=\frac{x_{int}^2}{DIR_{HI}}\exp\left(-\frac{IR_{HI}}{x_{int}}\right).
\label{eq:tau2}
\end{equation}

\begin{figure}[h]
\includegraphics[width=5in,height=5in,keepaspectratio]{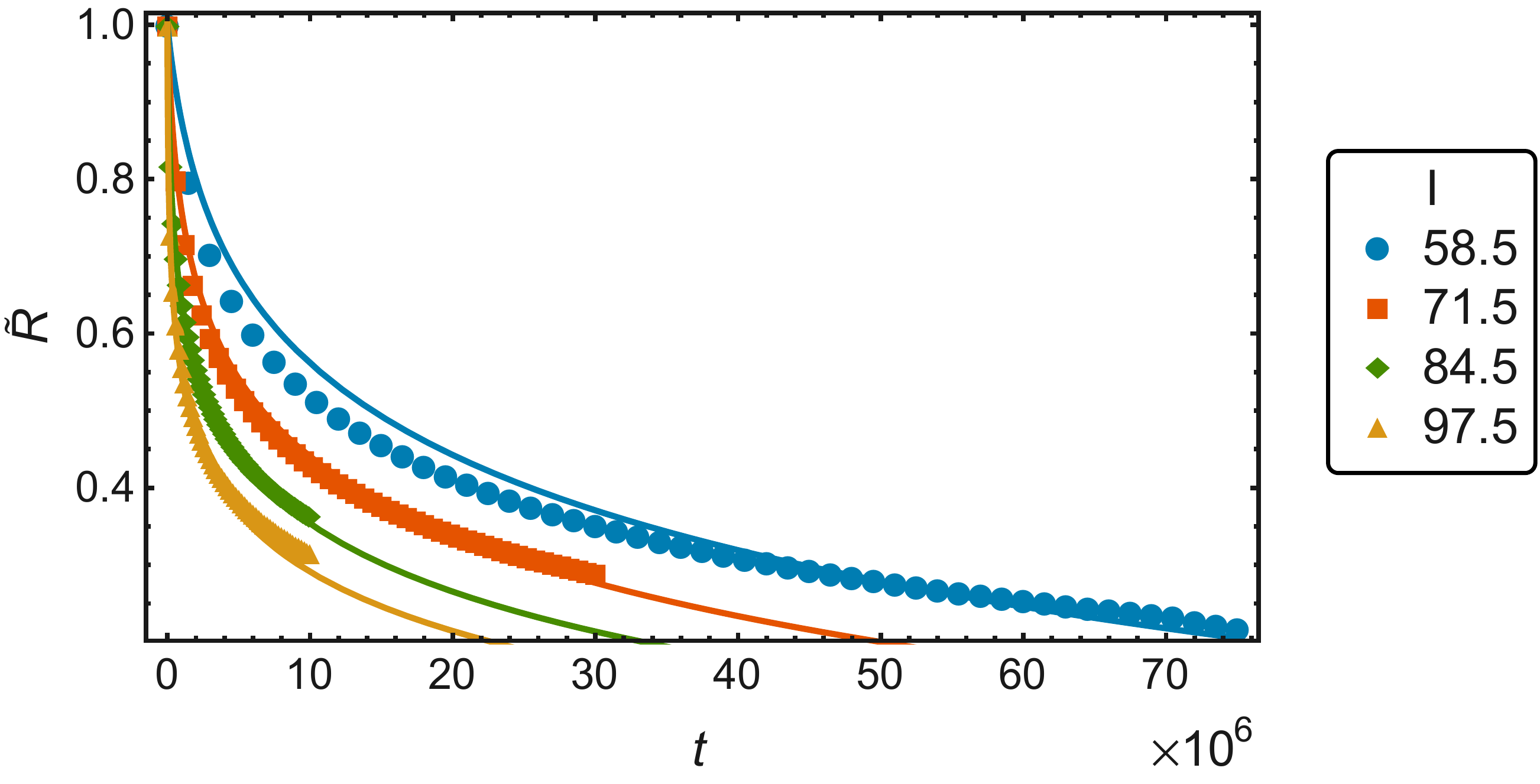}\caption{Fitting results for the evolution of the normalized resistance according to the Eq.~(\ref{eq:approximate solution}).}
\label{FittRbar}

\end{figure}

In Fig.S\ref{FittRbar} we use this approximate expression to fit the $R(t)$ simulation data. Comparison with the previous fit done with
Eq.\ref{eq:R(t)} and shown in Fig.S\ref{fig:R-fit} allows us to check that this approximate solution is relatively accurate within the time domain we are interested in for fast switching devices.

Now, using Eq.~(\ref{eq:approximate solution}), we can show that the $I$-dependent family curves $R(t)$
should obey scaling. 
We first consider the normalized resistive change $\delta R(t)$ defined as
\begin{eqnarray}
\delta R(t) & = & \frac{R(t) -R(\tau_{0})} {R_{HI}-R\left(\tau_{0}\right)}
\label{eq:scaling1}
\end{eqnarray}
where  $\tau_0$ is some yet unspecified time (and we drop the $*$ from $t^*$).   
Then, replacing with Eq.\ref{eq:approximate solution} we have,
\begin{eqnarray}
\delta R\left(t\right) & = & \frac{R_{HI} - \frac{x_{int}}{I} \ln(1+ \frac{t}{\tau_2}) - R_{HI} + \frac{x_{int}}{I} \ln(1+ \frac{\tau_0}{\tau_2})}
{R_{HI} - R_{HI} + \frac{x_{int}}{I} \ln(1+ \frac{\tau_0}{\tau_2})}
\end{eqnarray}
And rescaling the time by $\tau_0$,
\begin{eqnarray}
\delta R\left(t/\tau_0\right) & = & 1 - \frac{\ln(1+ \frac{\tau_0}{\tau_2}\frac{t}{\tau_0})} {\ln(1+ \frac{\tau_0}{\tau_2})}
\label{eq:scaling2}
\end{eqnarray}

Notice that with the natural choice of simply setting $\tau_0$ as $\tau_2$, 
we obtain the scaling form that we presented in the main text.

However, in the experiment we do not know, a priori, how to determine the characteristic time 
scale $\tau_2$, which is a strong function of the current $I$ and other material parameters. 
So we adopt the following strategy. We use $\tau_0$
as a free scaling parameter, one for each $I$-dependent curve $R(t)$, that we rescale according to 
Eq.\ref{eq:scaling2}.  We vary the set of values $\tau_0[I]$ until we obtain
a collapse of all the experimental and the simulation curves. 
The results of the successful collapse are shown in Fig.4 of the main text. In our experience, the collapse
is unique. 
A crucial point now is that the collapsed set of curves could be fitted with the expression
\begin{equation}
F(t)=1-\frac{\ln\left(1+ct\right)}{\ln\left(1+c\right)}.
\label{eq:scaling3}
\end{equation}
with $c$ a {\it current independent} constant. We find the values
$c=15.2$ for the experimental data and $c=29.4$ for the simulation ones. 
Hence, in regard of Eqs.\ref{eq:scaling2} and \ref{eq:scaling3} we observe that the constant $c$ is
nothing but the ratio between the analytically established characteristic time $\tau_2$ and the
empirically determined $\tau_0$, which are simply proportional to one another.

\begin{figure}[h]
\includegraphics[width=5in,height=5in,keepaspectratio]{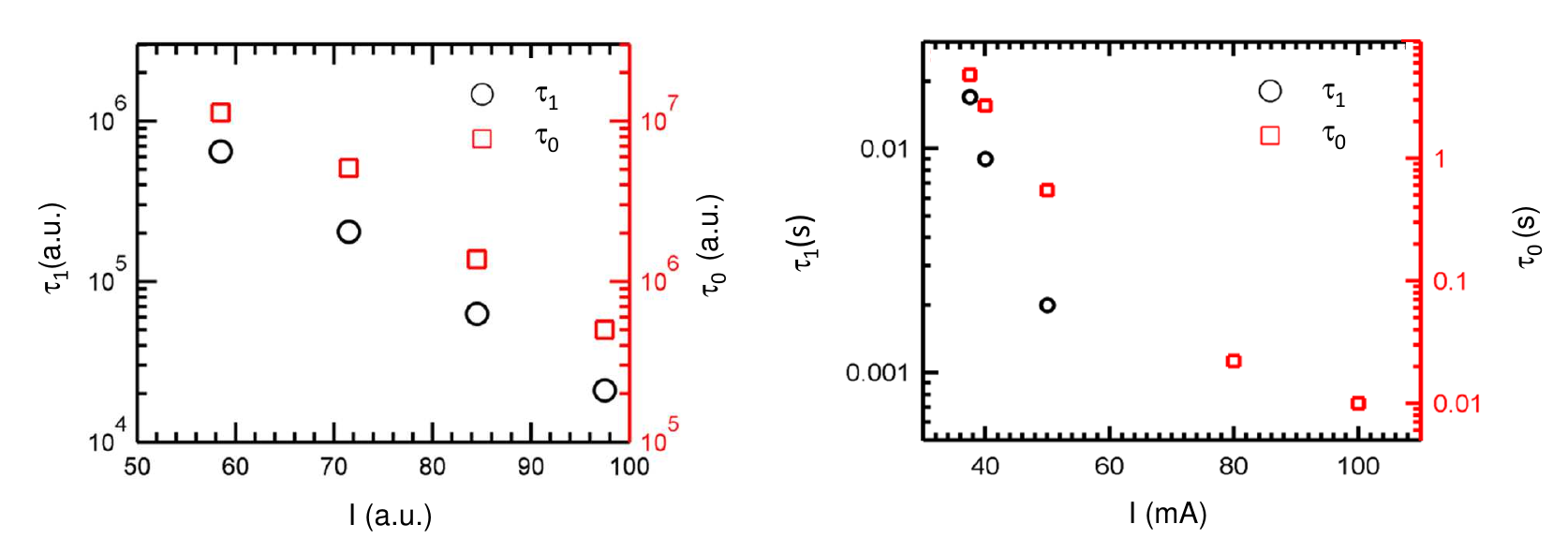}
\caption{Evolution of the characteristic times for different applied currents for both simulations (left panel) and 
experiments (right panel). It can be seen that both times follow an 
approximate exponential dependence with $I$, and that there exist a relative 
proportionality between them as predicted by our analysis.}
\label{Taustogether}

\end{figure}

In Fig.S\ref{Taustogether} we plot the dependence of the scaling time as a function of the applied
current $\tau_0(I)$.
In the case of the simulation results, the data follow the same exponentially decreasing 
behaviour as deduced for $\tau_2(I)$ (see Eq.\ref{eq:tau2}). 
Moreover, we find the ratio $\tau_0(I)/\tau_2(I)$ in good agreement with the constant $c=29.4$, which
validates our practical scaling procedure for the determination of the characteristic switching time in
the experimental case.

Under the same set of approximations that we have assumed in this section, plus the additional one
of neglecting the effect of the background distribution of vacancies, one may  also derive an
explicit expression for the characteristic time $\tau_1$.
We start from Eq.~(\ref{eq:shock wave velocity}), and similarly as before, adopt the 
approximation $\sinh\left(IR/x_{int}\right)\approx\frac{1}{2}\exp\left(IR/x_{int}\right)$. Then, making 
the substitution $y^{\prime} = 1/y$ in Eq.~(\ref{eq:t-x simulation}), and assuming that the integral is dominated 
by the exponential factor (i.e. neglecting the change in lower order factors), the integral has an analytical solution, 
and we get for the case with no background vacancies (i.e. $1/\beta =0$)
 \begin{equation}
 \tau_1 \approx C(t_0,x_0) + \frac{x_{int}^2}{DIR_{HI}} \exp\left(-\frac{IR_{HI}}{x_{int}}\right).
 \end{equation}
 
 where the first term is the integration constant determined by the initial conditions $t_0$  and $x_0$. If this constant can be neglected (for example for fast forming shock waves) then we have the surprising result that $\tau_1 = \tau_2$. 

In the case of the model simulations, the results displayed in Fig.S\ref{Taustogether} shows that, in fact, both characteristic times
have the same $I$-dependent exponentially decaying behaviour. Moreover, the ratio of the two characteristic times
is approximately 25, which is close to the constant $c=29.4$ quoted before, therefore our simulations
also validate the equality between the two characteristic times $\tau_1$ and $\tau_2$ predicted by the shock wave scenario.

As may be expected, the experimental situation is only qualitatively consistent with the previous discussion. As seen
in Fig.S\ref{Taustogether}, for the experimentally determined characteristic times, we observe that they have
approximately a similar dependence with the applied current, however, it is less clear if the equality between them also holds.

We show and compare the evolution of both characteristic times under different applied current in Fig.S\ref{Taustogether} for both simulations and experiments, observing a good agreement with the analysis presented here.

\section{Experimental details}

For the experimental validation of our theory, the RS phenomenon was studied in a bulk La$_{0.325}$Pr$_{0.300}$Ca$_{0.375}$MnO$_3$ (LPCMO) polycrystallline sample with hand painted mm seized Ag contacts. 

\begin{figure}[h!]
  \includegraphics[width=0.5\textwidth]{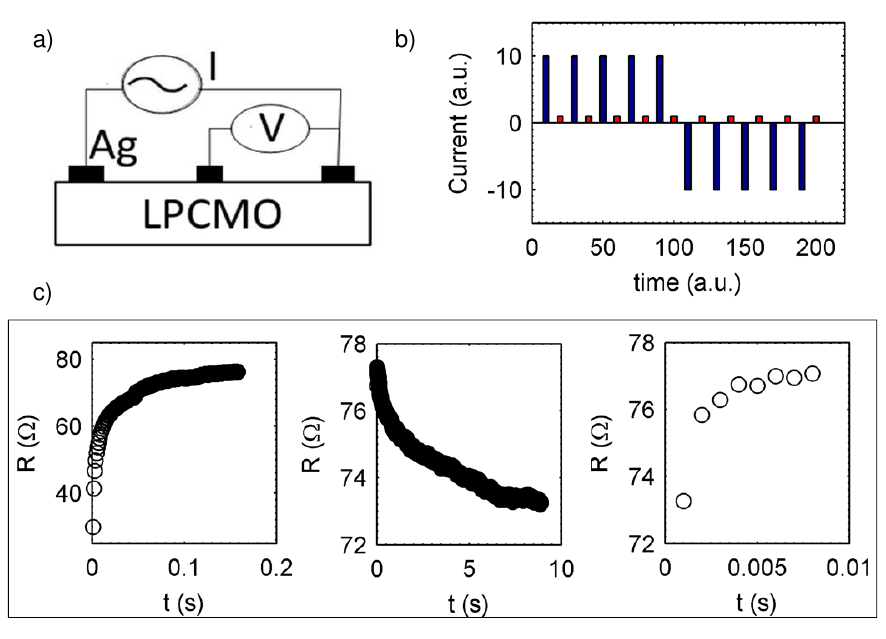}
  \caption{ (a) Diagram of the experimental set up. (b) Sketch of the pulsed current protocol used in the experiments. A hight current Write pulse (blue) is followed by a low current Read pulse (red). The first pulse generates the RS while the second pulse measures the non-volatile resistance. (c) Rvst curves from a complete RS process. The system is first taken to a Hi resistance state under a current of -350mA (left panel). Then the Hi to Lo RS is measured under an applied current of 37.5mA (middle panel). Finally the system is taken back to a Hi resistance state with the application of a -350mA current.
   }
  
  \label{fig:Fig-Exp}
\end{figure}

To induce the RS effect an external constant current was used as the electric stimulus. To generate this current and to acquire the data a Keithley 2612 Sourcemeter was employed. The measurements were done using a 3 wire configuration in order to measure a single interface resistance (Fig. \ref{fig:Fig-Exp}.a). The Hi to Lo RS was studied for currents of different magnitude, as described in the main text.

The application of current is done following a pulsed protocol: a high current pulse (Write) is followed by a low current pulse (Read). A schematic diagram of the pulsed protocol is shown in (Fig. \ref{fig:Fig-Exp}.b. Each pulse lasts 1ms. Between two pulses there is an interval of about 0.5s, meant to reduce  possible heating effects. The time axis exhibited in the R vs t curves (both in Fig.~3 in the main text, and in Fig. \ref{fig:Fig-Exp}.c, below) is the effective time elapsed during the actual application of the Write pulses, i.e. disregarding both the 0.5 s timeout and the Reading elapsed time. 
The Write pulses possess enough strength to change the resistive state of the system, while the Read pulse (low current) measures the remnant (stable, non-volatile) resistance of the device without affecting it. In the experiments shown in main text, the Write pulses are in the order of the mA while the Read pulses are in the order of the $\mu$A.

The complete measurement process is shown in Fig. \ref{fig:Fig-Exp}.c. To obtain the initial Hi state, pulses of -350mA were applied ( Fig. \ref{fig:Fig-Exp}.c, left panel). Next, the desired  accumulation experiment is performed, by applying positive pulses of constant amplitude which decrease the resistance (Fig. \ref{fig:Fig-Exp}.c, middle panel). The initial Hi -resistance state is recovered by applying -350 mA pulses (Fig. \ref{fig:Fig-Exp}.c, right panel). In every case the initial Hi-resistance state obtained was of the same magnitude within a range of about 1$\%$.

\bibliographystyle{apsrev}